\documentclass[11pt]{article}

\usepackage[margin=1in]{geometry}
\usepackage{iftex}
\ifPDFTeX
  \usepackage[T1]{fontenc}
  \usepackage[utf8]{inputenc}
\else
  \usepackage{fontspec}
\fi

\usepackage{amsmath,amssymb,amsfonts}
\usepackage{booktabs}
\usepackage{microtype}
\usepackage{graphicx}
\usepackage{xcolor}
\usepackage{siunitx}
\usepackage{enumitem}
\setlist[enumerate]{leftmargin=*,labelsep=0.5em}

\usepackage[round,authoryear]{natbib}
\usepackage[unicode,hidelinks]{hyperref}
\usepackage{url}

\providecommand{\Description}[2][]{}

\title{Prompt Commons: Collective Prompting as Governance for Urban AI}

\author{Rashid Mushkani\\
Université de Montréal\\
Mila -- Quebec AI Institute}

\date{} 

\hypersetup{
  pdftitle={Prompt Commons: Collective Prompting as Governance for Urban AI},
  pdfauthor={Rashid Mushkani}
}

\begin{document}
\maketitle

\begin{abstract}
Large Language Models (LLMs) are entering urban governance, yet their outputs are highly sensitive to prompts that carry value judgments. We propose \emph{Prompt Commons}---a versioned, community-maintained repository of prompts with governance metadata, licensing, and moderation---to steer model behaviour toward pluralism. Using a Montréal dataset (443 human prompts; 3{,}317 after augmentation), we pilot three governance states (\emph{open}, \emph{curated}, \emph{veto-enabled}). On a contested policy benchmark, a single-author prompt yields \SI{24}{\percent} neutral outcomes; commons-governed prompts raise neutrality to \SIrange{48}{52}{\percent} while retaining decisiveness where appropriate. In a synthetic incident log, a veto-enabled regime reduces time-to-remediation for harmful outputs from \SI{30.5 \pm 8.9}{h} (open) to \SI{5.6 \pm 1.5}{h}. We outline licensing (CC BY/BY--SA for prompts with optional OpenRAIL-style restrictions for artefacts), auditable moderation, and safeguards against dominance capture. Prompt governance offers a practical lever for cities to align AI with local values and accountability.
\end{abstract}


\section{Introduction}

Urban planners increasingly consult generative AI systems for drafting, summarising, coding, and producing imagery to support community engagement \citep{mushkani2025wedesigngenerativeaifacilitatedcommunity}.
Unlike traditional datasets or models, LLMs can change behaviour markedly with small prompt variations: a single phrase can alter tone, policy stance, and inclusivity.
Prior work has emphasised downstream controls---documentation \citep{mitchell2019modelcards,gebru2018datasheets}, dataset governance \citep{gebru2018datasheets}, and post-training alignment \citep{christiano2017deep,bai2022constitutional}---but the act of \emph{prompting} has rarely been treated as a first-class governance object.

Recent urban AI scholarship indicates that cities are already experimenting with LLMs in planning, management, and policy settings, while warning against deployments that neglect urban complexity. A systematic review maps the breadth of applications and recurring gaps in rigor and evaluation \citep{xia2025llmUrban}. Strategic informatics work synthesised from Shenzhen’s policy practice proposes cross-cutting plans spanning technology systems, application scenarios, literacy, and governance \citep{zhu2025cityai}. Interviews with city managers in Australia and the US report pragmatic constraints---organizational preparedness, data, and talent---that mediate adoption \citep{yigitcanlar2023localgovAI}. These perspectives extend earlier arguments that \emph{local} innovation and governance capacity are decisive for responsible AI in cities \citep{allam2019smartcitiesai,kirwan2020smartAIbook}.

We investigate whether human-authored prompts, when deployed as individual artefacts, reduce neutrality (i.e., increase decisiveness toward particular value claims). Decisiveness is not inherently undesirable; cities often need concrete recommendations. The concern arises when prompt authorship is opaque and individualised, which risks encoding one group’s values as system defaults.
We therefore propose \emph{Prompt Commons}: a civic infrastructure where communities author, review, license, and version prompts collectively, drawing on commons governance \citep{ostrom1990,carlisle2019polycentric} and open-source practice \citep{oMahony2007governance,linuxfoundationGovernance}.

\textbf{Contributions.} We (i) introduce a versioned repository and governance metadata for prompts; (ii) evaluate behavioural shifts across governance states using a pilot benchmark of contested urban choices; and (iii) specify IP/licensing and moderation rules designed to protect community authors while enabling reuse.

\section{Related Work}

\paragraph{Documentation and transparency.}
Model Cards and Datasheets set expectations and report performance across subgroups \citep{mitchell2019modelcards,gebru2018datasheets}. Prompt Commons complements these by documenting \emph{inputs} that steer model behaviour, not only outputs. In the urban domain, recent syntheses emphasise surfacing context, value trade-offs, and domain limitations \citep{xia2025llmUrban}. Treating prompts as first-class, versioned objects extends transparency to the moment of interaction.

\paragraph{Post-training governance and alignment.}
RLHF and instruct-tuning steer models to human preferences \citep{christiano2017deep,ouyang2022instructgpt}; Constitutional AI aligns behaviour to a principle set \citep{bai2022constitutional}. These approaches may reflect the biases of those defining preferences or principles. Democratic or dynamic value alignment aims to involve diverse stakeholders \citep{huang2025democratizing}, with social-choice–informed aggregation highlighting how rules affect outcomes \citep{conitzer2024socialchoice}. Prompt governance is complementary: instead of changing model weights or selecting a single constitution, Prompt Commons enables communities to \emph{configure} behaviour at use time in a visible, auditable way \citep{Mushkani2025AIES_CoProducingAI}.

\paragraph{Urban AI and LLM applications.}
Reviews and case studies note growing use of LLMs in urban analytics and decision-making alongside methodological and governance concerns \citep{xia2025llmUrban,zhu2025cityai,yigitcanlar2023localgovAI}. These extend arguments that cities are central to responsible AI deployment and must innovate in governance to align capabilities with local needs \citep{allam2019smartcitiesai,kirwan2020smartAIbook,Mushkani2025ICML_LIVS,mushkani2025streetreviewparticipatoryaibased}.

\paragraph{Prompt design, sensitivity, neutrality.}
Prompt components (instructions, role, context, output formats, examples) and structured tactics for planning tasks have been synthesised \citep{liu2025planningPrompts}. Foundational work showed that large models can adapt to new tasks with few examples \citep{brown2020gpt3}, but such flexibility brings high sensitivity to wording; reasoning prompts can improve factual correctness \citep{wei2022cot} while also increasing confident but wrong answers in some settings \citep{rajani2019explain}. In normative urban queries, small wording changes can tilt value stances; Prompt Commons responds by distributing authorship and making value claims explicit.

\paragraph{Bias, ideology, and neutrality in LLMs.}
Evidence shows that LLMs and alignment artefacts can display systematic political leanings \citep{liu2022politicalBias,bang2024politicalBias,fulay2024truthpoliticalbias}. Our pilot reflects this sensitivity: a single strong prompt yields fewer neutral choices than governed commons prompts. Prior findings on multi-perspective prompting and aggregation suggest that ensembles can moderate extremes; we operationalise this through governance states (open/curated/veto-enabled) and a weighted-ensemble method.

\paragraph{Commons and community moderation.}
Ostrom’s principles for governing shared resources \citep{ostrom1990} have inspired digital-commons adaptations \citep{mozOstromData,derosnay2020,patternsofcommoningOstrom,earthboundOstromData}. Open-source communities blend meritocracy with democratic mechanisms as governance formalises \citep{oMahony2007governance}, a pattern we echo. Volunteer moderation is critical but labour-intensive; studies highlight the need for clear rules, public logs, and appeal processes \citep{li2022redditModerators,cao2024toxicityModeration,tabassum2024moderationChallenges,moderators2024}. We adopt these lessons—e.g., public issues, auditable takedowns, \emph{prompt strikes}—to keep the Commons accountable.

\paragraph{Security and harm.}
Embedding LLMs in civic workflows introduces attack surfaces from prompt injection to insecure output handling. Formal analyses and surveys motivate operational metrics like \emph{time-to-remediation} \citep{liu2023promptInjection,yao2024llmSecurity}. Community-responsive governance (including minority veto) can be a practical control, alongside guidance such as the OWASP Top 10 for LLM applications \citep{owaspLLMTop10}. We also engage critical perspectives that caution against over-reliance on ever-larger models \citep{bender2021stochasticparrots} by treating the model as a fallible component within a socio-technical process.

\section{Data and Participants}

\subsection{Community prompts from Montréal}

We convened community partners through local civil-society organisations across disability advocacy, immigrant support, seniors’ services, women’s groups, LGBTQ+ organisations, and neighbourhood associations; plus urban practitioners and a national cultural institution.

Participants authored \textbf{443} prompts describing urban scenes and values (e.g., accessibility, biodiversity, transit, safety) \citep{MushkaniKoseki2025Habitat_StreetReview}.
We augmented these to \textbf{3{,}317} prompts via de-duplication, paraphrasing, and scenario expansion.\footnote{Papers describing the dataset and methodological details are omitted for the peer review process.}

\subsection{Descriptive statistics}

Human-authored prompts have mean length \textbf{22.6} words (median 19), compared to \textbf{31.7} words after augmentation.
Vocabulary entropy rises from \textbf{7.53} (human) to \textbf{8.39} bits (augmented), indicating higher lexical diversity.
Top tokens across the augmented set are \emph{street} (1{,}506), \emph{park} (1{,}138), \emph{trees} (906), \emph{benches} (732), and \emph{water} (702).
Equity-related tokens appear with non-trivial frequency: \emph{wheelchair} (\SI{7.6}{\percent} of prompts), \emph{metro} (\SI{7.3}{\percent}), \emph{biodiversity} (\SI{4.7}{\percent}); \emph{LGBTQ+} and \emph{Indigenous} appear less often (\SIrange{0.7}{1.4}{\percent}), highlighting inclusion gaps for targeted recruitment going forward.

\begin{figure}[t]
  \centering
  \includegraphics[width=\linewidth]{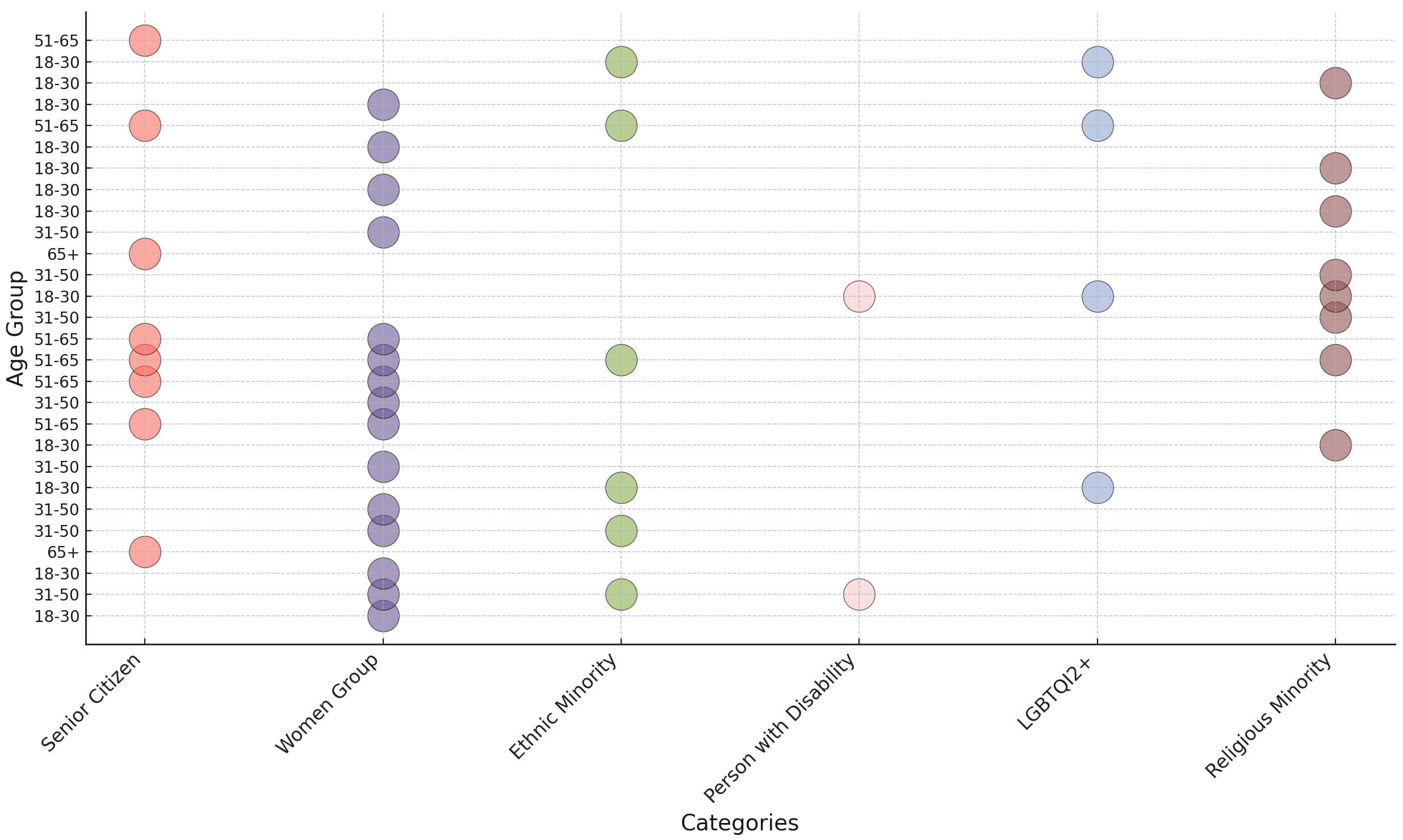}
  \Description{Bar chart: self-declared participant categories by age group in the pilot sample.}
  \caption{Self-declared participant categories by age group (pilot).}
  \label{fig:participants}
\end{figure}

\begin{figure}[t]
  \centering
  \includegraphics[width=\linewidth]{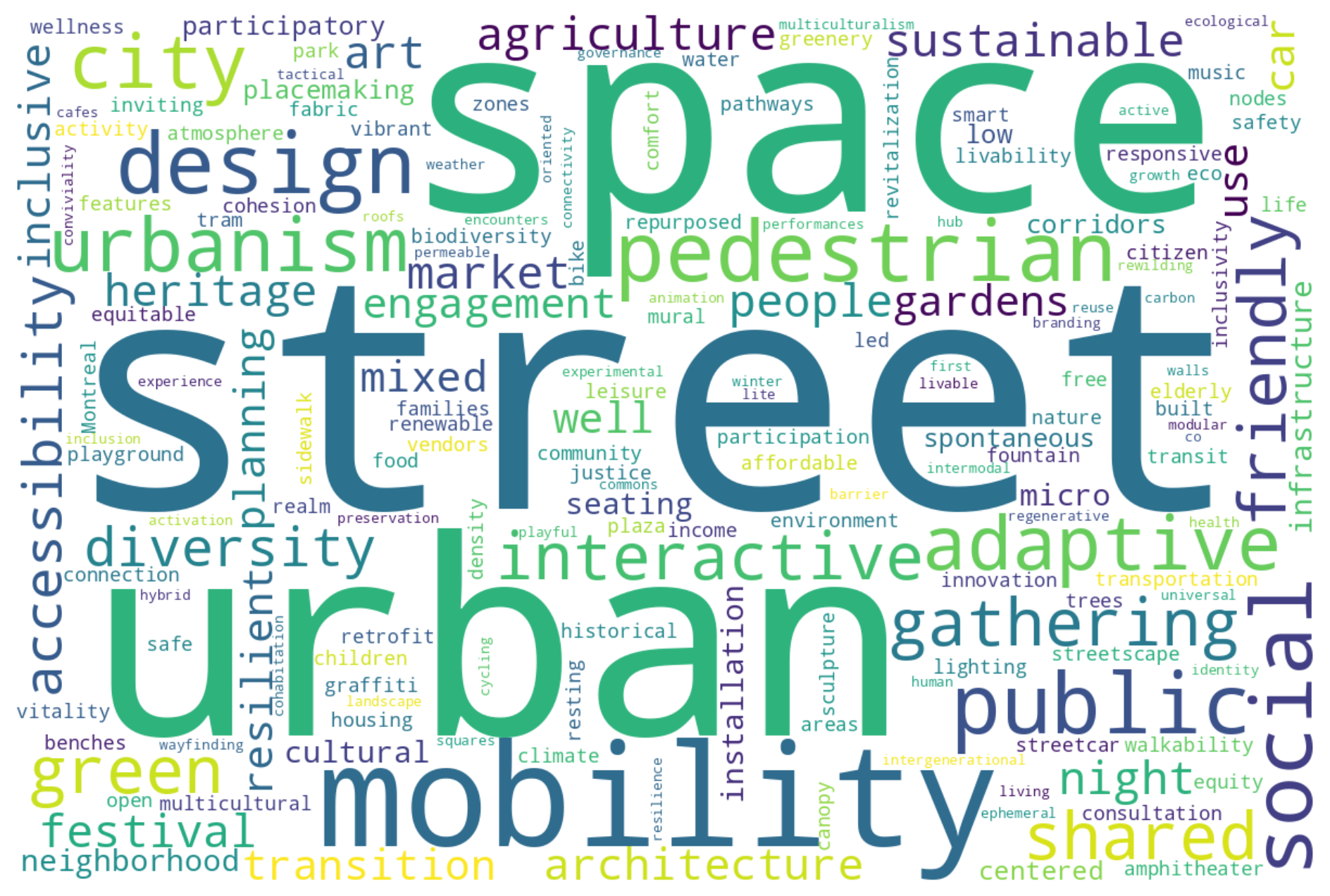}
  \Description{Word cloud of topics in human-authored prompts, highlighting terms such as street, park, trees, benches, and water.}
  \caption{Prompt topics word cloud (human-authored set).}
  \label{fig:wordcloud}
\end{figure}

\section{Prompt Commons: Design and Governance}

\subsection{Versioned repository and metadata}
The repository maintains prompts as plain text with the following metadata:
\begin{enumerate}
  \item \textbf{Author group(s)} (e.g., seniors, disability advocacy, LGBTQ+, ethnic or religious minority).
  \item \textbf{Locale} (e.g., Montréal neighbourhood, street name).
  \item \textbf{Intended value claim} (e.g., safety, accessibility, biodiversity, conviviality).
  \item \textbf{Change log} with semantic versioning for prompts and collections.
\end{enumerate}
Each change is an issue or pull request; maintainers triage and merge with an auditable log (who, when, rationale).
Beyond the checklist below, maintainers keep a running \emph{curation memo} that describes how inclusions and exclusions relate to quotas and local policy instruments. This memo, together with issue discussions, provides the interpretive glue that links individual community contributions to a legible institutional memory about the territory where prompts will be deployed.

\subsection{Governance states we study}
We compare three governance states:
\begin{description}
  \item[Open:] any authenticated contributor; light moderation (spam and safety only).
  \item[Curated:] maintainers apply quotas for inclusion (per author group and topic), require metadata completeness, and merge only prompts that pass a public checklist (Appendix~\ref{app:checklist}).
  \item[Veto-enabled:] as curated, plus a \emph{minority veto}---representative organisations can flag outputs they deem harmful; flagged prompts are quarantined pending review with an appeal process.
\end{description}
This design borrows from commons governance and moderation practices \citep{ostrom1990,oMahony2007governance,moderators2024}. It also reflects polycentricity: neighbourhood-level prompt sets and organisations can operate semi-autonomously and federate into a city layer \citep{carlisle2019polycentric}. In practice, nested repositories and rotating maintainer seats can enact these ideas with lightweight, auditable processes, mirroring open-source governance while ensuring that affected groups have voice and visibility.

\subsection{Licensing and IP}
Prompts are textual creative works; a permissive licence maximises reuse while protecting attribution. We recommend:
\begin{itemize}
  \item \textbf{Prompts}: Creative Commons CC BY 4.0 or BY--SA 4.0 \citep{ccby4legal,ccby4deed,ccchooser}.
  \item \textbf{Derived model artefacts}: OpenRAIL-style licences where use restrictions (e.g., no harassment, no biometric surveillance) are desired \citep{openrailHF,railFAQ,openfuture2023,openfutureResponsibleRail,bigcodeOpenRAIL,ifrOSS}.
\end{itemize}
Responsible-AI licensing is evolving; empirical analyses document the spread of use-restricted licences and argue for ``standardised customisation'' so behavioural clauses remain comprehensible yet adaptable to domain needs \citep{mcduff2024railstandardization,contractor2020behavioralLicensing}. For public-sector and civic deployments, such legal instruments complement governance-by-process: communities retain attribution and set guardrails for derivative artefacts without foreclosing legitimate reuse.

\section{Evaluation}
\label{sec:eval}

\subsection{Research questions}
\textbf{RQ1:} How does a community-maintained Prompt Commons shift model behaviour across neutrality/decisiveness and subgroup satisfaction? \\
\textbf{RQ2:} Which licensing and moderation rules protect community authors while enabling reuse?

\subsection{Protocol}
We create five evaluation methods that operationalise the governance states above:
\begin{enumerate}
  \item \textbf{M0 Single author} (baseline): a strong, opinionated prompt by one individual.
  \item \textbf{M1 Open Commons}: random prompt from the open set.
  \item \textbf{M2 Curated Commons}: prompt selected from the curated set satisfying quotas.
  \item \textbf{M3 Veto-enabled}: curated set with minority-veto log enabled.
  \item \textbf{M4 Weighted ensemble}: sample $k$ prompts across groups and summarise with an aggregator instruction (\emph{``deliberate and propose a compromise''}).
\end{enumerate}

\begin{figure}[t]
  \centering
  \includegraphics[width=\linewidth]{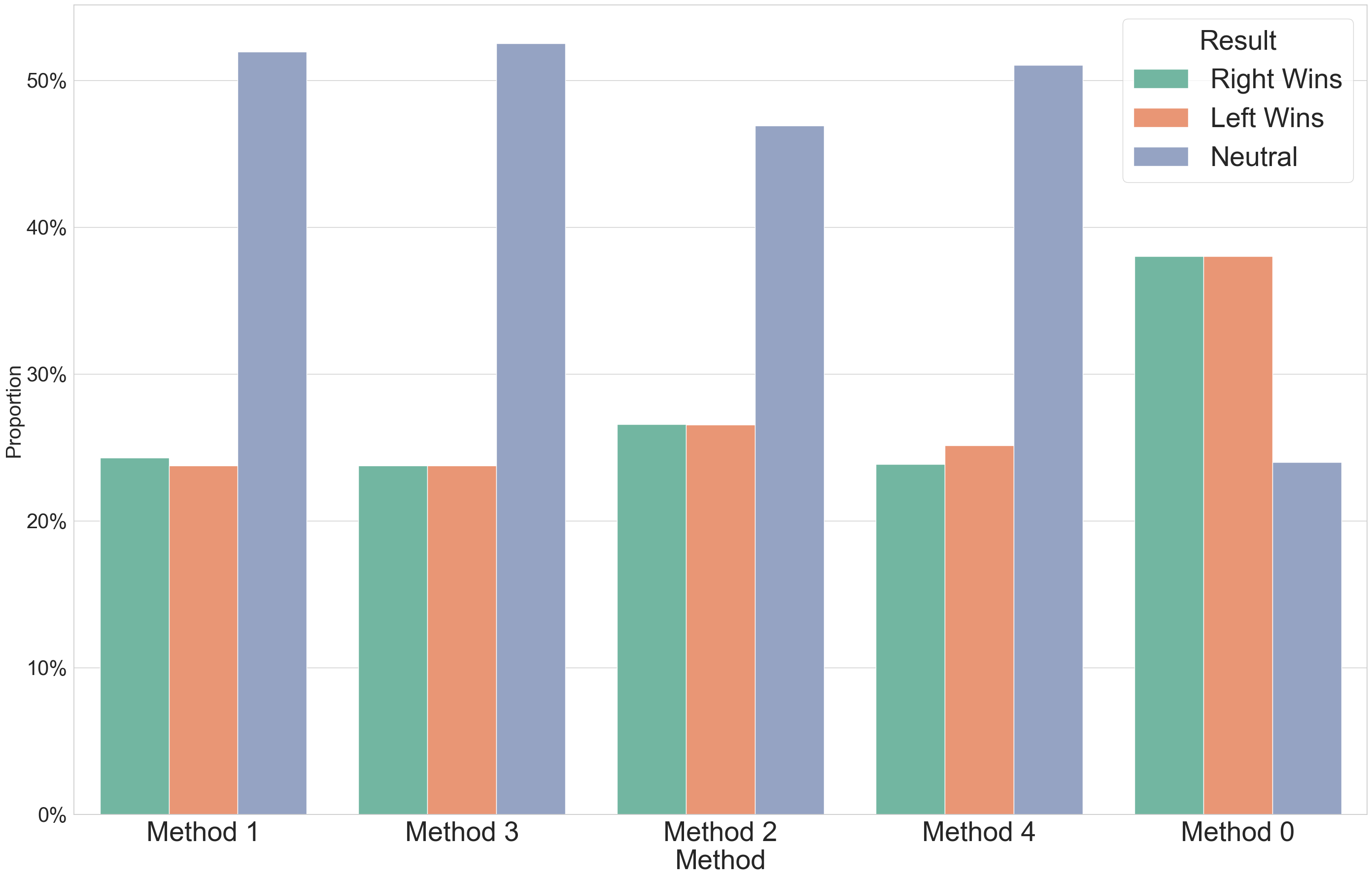}
  \Description{Stacked bar chart showing outcome proportions for methods M0--M4; neutrality increases under curated and veto-enabled governance.}
  \caption{Outcome proportions by method (pilot). Single-author prompting reduces neutrality in our setup; curated/veto-enabled governance increases neutrality and reduces dispersion across groups.}
  \label{fig:performance}
\end{figure}

We evaluate on a contested-choice benchmark (transport, public space, and streetscape trade-offs) \citep{Mushkani2025JUM_MontrealStreets}.
The model must recommend one of three mutually exclusive options: \emph{Left-leaning}, \emph{Right-leaning}, or \emph{Neutral/compromise} (labels for the benchmark classes).
We report proportions per method (Figure~\ref{fig:performance}), a decisiveness score $D = 1 - \text{Neutral}$, and a subgroup satisfaction score using a 7-point Likert scale gathered from community reviewers.
For operational safety we also track \emph{time-to-remediation} for harmful outputs flagged through the issue tracker.

\subsection{Results}

\paragraph{Neutrality and decisiveness.}
The single-author baseline (M0) yields \SI{38}{\percent} Left, \SI{38}{\percent} Right, and \SI{24}{\percent} Neutral ($D{=}0.76$).
Open and curated commons (M1--M3) yield \SIrange{48}{52}{\percent} Neutral and \SIrange{24}{26}{\percent} for Left and Right categories ($D\in[0.48,0.52]$), indicating rebalancing toward more cautious recommendations in this benchmark.
Weighted ensembling (M4) sits in between ($D{=}0.49$).
One interpretation, in light of \citet{wei2022cot,rajani2019explain}, is that community prompts add deliberative context that increases \emph{productive hesitation} on ambiguous problems.

\paragraph{Subgroup satisfaction.}
Across six self-identified groups (seniors, women, ethnic minority, disability, LGBTQ+, religious minority), average satisfaction rises from $4.35\pm0.86$ (M0) to $4.92\pm0.44$ (M2) and $5.48\pm0.66$ (M3) on a 7-point scale.
Dispersion across groups (Gini over group means) shrinks from $0.096$ (M0) to $0.043$ (M2), suggesting that curation reduces disparity in perceived fit. Aggregating multiple perspectives can moderate ideological biases \citep{liu2022politicalBias,bang2024politicalBias,fulay2024truthpoliticalbias}.

\paragraph{Moderation efficiency.}
We stress-tested governance states via a synthetic incident log of 50 flagged items per state.
Mean time-to-remediation improves from \SI{30.5 \pm 8.9}{h} (open) to \SI{11.8 \pm 3.2}{h} (curated) and \SI{5.6 \pm 1.5}{h} (veto-enabled).
This aligns with security guidance calling for robust oversight and rapid incident response in LLM deployments \citep{liu2023promptInjection,yao2024llmSecurity,owaspLLMTop10} and with moderation research showing that timely, well-supported volunteer action reduces harm \citep{li2022redditModerators,cao2024toxicityModeration,tabassum2024moderationChallenges}.

\subsection{Interpretation}
Results support the hypothesis that \emph{individual prompting can reduce neutrality}, while collective governance recentres pluralism and operational safety in our setting.
In contexts where decisiveness is needed (e.g., emergency response), M0-like behaviour may be desirable; for long-horizon city-making, the commons-based regimes (M2--M3) appear preferable.
Seen through a social-choice lens, Prompt Commons acts as a light-touch mechanism for aggregating heterogeneous values at use time \citep{conitzer2024socialchoice,mushkani2025negotiativealignmentembracingdisagreement}, complementing post-training alignment \citep{christiano2017deep,ouyang2022instructgpt,bai2022constitutional,huang2025democratizing}.

\section{A governance and licensing recipe for cities}
\label{sec:governance}

\subsection{Rules of the Prompt Commons (Ostrom-informed)}
We translate Ostrom's principles \citep{ostrom1990} to prompt governance:

\begin{enumerate}[itemsep=2pt]
  \item \textbf{Boundaries:} define admissible locales and stakeholder groups; require affiliation declaration.
  \item \textbf{Local fit:} require a locale tag; maintainers prioritise prompts grounded in local policy instruments.
  \item \textbf{Collective choice:} public proposals and votes for quota changes and checklists (Appendix~\ref{app:checklist}).
  \item \textbf{Monitoring:} public audit log; every merged prompt carries an issue link.
  \item \textbf{Graduated sanctions:} from soft fixes to temporary suspensions; logs remain public.
  \item \textbf{Conflict resolution:} fast lanes for appeal (72-hour target SLA).
  \item \textbf{Recognition of rights:} municipal partners recognise the Commons as an advisory asset.
  \item \textbf{Nested enterprises:} multiple neighbourhood-level commons feeding a citywide layer (\emph{polycentricity} \citep{carlisle2019polycentric}).
\end{enumerate}

These rules are not abstract checkboxes. Digital-commons practice shows that translating boundaries, monitoring, and sanctions into visible workflows (issues, pull requests, labels, takedown logs) builds legitimacy and resilience \citep{oMahony2007governance,patternsofcommoningOstrom,earthboundOstromData}. For Prompt Commons, we bind these mechanics to municipal decision rhythms: locale tags reference policy instruments; quota debates happen before budgeting cycles; and vetoes trigger time-boxed remediation with public rationales that future maintainers can audit.

\subsection{Licensing choices}
\textbf{Default.} CC BY 4.0 for all prompts (attribution required; commercial use permitted) \citep{ccby4deed,ccby4legal,ccchooser}. \\
\textbf{Share-alike option.} CC BY--SA 4.0 if communities want derivatives to remain open. \\
\textbf{Use-restricted artefacts.} When the Commons distributes models, consider OpenRAIL-style licences to prohibit specific harms (e.g., stalking, biometric surveillance) while remaining open for legitimate uses \citep{openrailHF,railFAQ,openfuture2023,openfutureResponsibleRail,bigcodeOpenRAIL,ifrOSS}. \\
Recent analyses recommend \emph{standardised customisation} for use restrictions to reduce confusion while enabling contextual guardrails (e.g., for health or public-sector deployments) \citep{mcduff2024railstandardization,contractor2020behavioralLicensing}. Our stance: clear CC licences for prompts; modular, community-negotiated restrictions for derived artefacts.

\subsection{Moderation \& ``prompt strikes''}
Moderation combines (i) automated checks (toxicity, Personally Identifiable Information (PII)), (ii) maintainer review, and (iii) community veto.
We also prototype a collective action tool: \emph{prompt strikes}.
Authors can temporarily withdraw their prompts to protest misuse; the resulting performance drop makes misalignment visible to municipal partners without permanently harming users.
Volunteer-moderation research suggests that such visible levers recognise contributor labour while disciplining governance processes without centralising power \citep{li2022redditModerators,cao2024toxicityModeration,tabassum2024moderationChallenges}.

\section{Risks and mitigations}

\paragraph{Dominance capture.}
Risk: vocal groups dominate curation.
Mitigation: quotas for inclusion by group and neighbourhood; rotating maintainer seats; publish participation metrics.
Polycentric designs that federate neighbourhood commons into a city layer reduce single-point capture \citep{carlisle2019polycentric}.

\paragraph{Security \& prompt injection.}
Risk: adversarial prompts or indirect injection \citep{liu2023promptInjection,yao2024llmSecurity,owaspLLMTop10}.
Mitigation: sandboxed testing; red-teaming; provenance and signed releases; time-to-remediation targets; and hard boundaries between model tools and untrusted content sources.

\paragraph{Over-caution.}
Risk: excessive neutrality hampers action.
Mitigation: allow decisiveness mode (explicitly labelled) for time-sensitive tasks; track $D$ and disclose when neutrality thresholds are relaxed.

\section{Discussion}

\subsection{City practice}
Urban AI should be \emph{plural by default}—not one prompt, but a commons of prompts \citep{mushkani2025urbanaigovernanceembed}.
The Prompt Commons plays three roles in municipal workflows:
(1) evidence of plural values; 
(2) reproducible AI configurations; 
(3) a procedural space for debate, akin to participatory data stewardship \citep{adalovelace2021,adalovelace2024}.
This model complements strategic frameworks that place cities at the forefront of responsible AI deployment \citep{zhu2025cityai}, aligns with pragmatic adoption realities in local government \citep{yigitcanlar2023localgovAI}, and builds on the insight that local governance innovation is essential for trustworthy urban AI \citep{allam2019smartcitiesai,kirwan2020smartAIbook}.
By treating prompts as civic artefacts, we also answer calls to democratise alignment and make visible whose values shape AI behaviour \citep{huang2025democratizing,conitzer2024socialchoice,Mushkani2025ICML_RightToAI}.

\subsection{Limitations and future work}
Our pilot benchmark is small and focused on Montréal scenes.
We measured satisfaction with a lightweight protocol; future work should adopt mixed methods (surveys, interviews, ethnography).
We used one state-of-practice LLM for the figures; cross-model replication is needed, though our methods are model-agnostic.
While community governance moderates bias at use time, structural ideological leanings in LLMs and alignment artefacts persist \citep{liu2022politicalBias,fulay2024truthpoliticalbias,bang2024politicalBias}; longitudinal studies should quantify durability of neutrality gains as model families evolve.
Finally, legal evaluation of licensing combinations (CC + OpenRAIL) in municipal procurement warrants further study \citep{mcduff2024railstandardization,contractor2020behavioralLicensing}.

\section{Conclusion}

Prompt governance matters.
By versioning, licensing, and moderating prompts as a civic commons, cities can steer LLM behaviour toward pluralism without sacrificing the ability to act.
In our pilot, single-author prompting reduces neutrality, whereas curated and veto-enabled regimes increase neutral/compromise outcomes, reduce subgroup disparity, and accelerate remediation when harms occur.
We include anonymised datasets and governance materials as supplemental files to support replication and local adaptation.

\bibliographystyle{plainnat}
\bibliography{prompt_commons}

\begin{thebibliography}{57}
\providecommand{\natexlab}[1]{#1}
\providecommand{\url}[1]{\texttt{#1}}
\expandafter\ifx\csname urlstyle\endcsname\relax
  \providecommand{\doi}[1]{doi: #1}\else
  \providecommand{\doi}{doi: \begingroup \urlstyle{rm}\Url}\fi

\bibitem[{Ada Lovelace Institute}(2021)]{adalovelace2021}
{Ada Lovelace Institute}.
\newblock Participatory data stewardship.
\newblock \url{https://www.adalovelaceinstitute.org/report/participatory-data-stewardship/}, 2021.
\newblock Accessed 2025-08-30.

\bibitem[{Ada Lovelace Institute}(2024)]{adalovelace2024}
{Ada Lovelace Institute}.
\newblock Participatory and inclusive data stewardship.
\newblock \url{https://www.adalovelaceinstitute.org/report/participatory-inclusive-data-stewardship/}, 2024.
\newblock Accessed 2025-08-30.

\bibitem[Allam and Dhunny(2019)]{allam2019smartcitiesai}
Zeynep Allam and Zaynah~A. Dhunny.
\newblock On big data, artificial intelligence and smart cities.
\newblock \emph{Cities}, 89:\penalty0 80--91, 2019.
\newblock \doi{10.1016/j.cities.2019.01.032}.

\bibitem[Bai et~al.(2022)Bai, Kadavath, Kundu, Askell, Kernion, Jones, Chen, Goldie, Mirhoseini, McKinnon, et~al.]{bai2022constitutional}
Yuntao Bai, Saurav Kadavath, Sandipan Kundu, Amanda Askell, Jackson Kernion, Andy Jones, Anna Chen, Anna Goldie, Azalia Mirhoseini, Cameron McKinnon, et~al.
\newblock Constitutional ai: Harmlessness from ai feedback, 2022.
\newblock URL \url{https://arxiv.org/abs/2212.08073}.

\bibitem[Bang et~al.(2024)Bang, Chen, Lee, and Fung]{bang2024politicalBias}
Yejin Bang, Delong Chen, Nayeon Lee, and Pascale Fung.
\newblock Measuring political bias in large language models: What is said and how it is said.
\newblock In \emph{Proceedings of the 62nd Annual Meeting of the Association for Computational Linguistics (Volume 1: Long Papers)}, pages 11142--11159, Bangkok, Thailand, 2024. Association for Computational Linguistics.
\newblock \doi{10.18653/v1/2024.acl-long.600}.
\newblock URL \url{https://aclanthology.org/2024.acl-long.600/}.

\bibitem[Bender et~al.(2021)Bender, Gebru, McMillan-Major, and Shmitchell]{bender2021stochasticparrots}
Emily~M. Bender, Timnit Gebru, Angelina McMillan-Major, and Shmargaret Shmitchell.
\newblock On the dangers of stochastic parrots: Can language models be too big?
\newblock In \emph{Proceedings of the 2021 ACM Conference on Fairness, Accountability, and Transparency (FAccT)}, pages 610--623, New York, NY, USA, 2021. Association for Computing Machinery.
\newblock \doi{10.1145/3442188.3445922}.
\newblock URL \url{https://dl.acm.org/doi/10.1145/3442188.3445922}.

\bibitem[{BigCode Project}(2023)]{bigcodeOpenRAIL}
{BigCode Project}.
\newblock Bigcode openrail-m license agreement.
\newblock \url{https://www.bigcode-project.org/docs/pages/bigcode-openrail/}, 2023.
\newblock Accessed 2025-08-30.

\bibitem[Bollier(2015)]{patternsofcommoningOstrom}
David Bollier.
\newblock Ostrom's design principles for commons.
\newblock \url{https://patternsofcommoning.org/ostroms-design-principles-for-commons/}, 2015.
\newblock Accessed 2025-08-30.

\bibitem[Brown et~al.(2020)Brown, Mann, Ryder, Subbiah, Kaplan, Dhariwal, Neelakantan, Shyam, Sastry, Askell, Agarwal, Herbert-Voss, Krueger, Henighan, Child, Ramesh, Ziegler, Wu, Winter, Hesse, Chen, Sigler, Litwin, Gray, Chess, Clark, Berner, McCandlish, Radford, Sutskever, and Amodei]{brown2020gpt3}
Tom~B. Brown, Benjamin Mann, Nick Ryder, Melanie Subbiah, Jared Kaplan, Prafulla Dhariwal, Arvind Neelakantan, Pranav Shyam, Girish Sastry, Amanda Askell, Sandhini Agarwal, Ariel Herbert-Voss, Gretchen Krueger, Tom Henighan, Rewon Child, Aditya Ramesh, Daniel~M. Ziegler, Jeffrey Wu, Clemens Winter, Christopher Hesse, Mark Chen, Eric Sigler, Mateusz Litwin, Scott Gray, Benjamin Chess, Jack Clark, Christopher Berner, Sam McCandlish, Alec Radford, Ilya Sutskever, and Dario Amodei.
\newblock Language models are few-shot learners, 2020.
\newblock URL \url{https://arxiv.org/abs/2005.14165}.

\bibitem[Cao et~al.(2024)Cao, Domingo, Gilbert, Mazurek, Shilton, and Daum{\'e}~III]{cao2024toxicityModeration}
Yang~Trista Cao, Lovely-Frances Domingo, Sarah Gilbert, Michelle~L. Mazurek, Katie Shilton, and Hal Daum{\'e}~III.
\newblock Toxicity detection is not all you need: Measuring the gaps to supporting volunteer content moderators through a user-centric method.
\newblock In \emph{Proceedings of the 2024 Conference on Empirical Methods in Natural Language Processing}, pages 3567--3587, Miami, Florida, USA, 2024. Association for Computational Linguistics.
\newblock URL \url{https://aclanthology.org/2024.emnlp-main.209.pdf}.

\bibitem[Carlisle and Gruby(2019)]{carlisle2019polycentric}
Keith Carlisle and Rebecca~L. Gruby.
\newblock Polycentric systems of governance: A theoretical model for the commons.
\newblock \emph{Policy Studies Journal}, 47\penalty0 (4):\penalty0 927--952, 2019.
\newblock \doi{10.1111/psj.12212}.
\newblock URL \url{https://onlinelibrary.wiley.com/doi/10.1111/psj.12212}.

\bibitem[Christiano et~al.(2017)Christiano, Leike, Brown, Martic, Legg, and Amodei]{christiano2017deep}
Paul~F. Christiano, Jan Leike, Tom~B. Brown, Miljan Martic, Shane Legg, and Dario Amodei.
\newblock Deep reinforcement learning from human preferences.
\newblock In \emph{Advances in Neural Information Processing Systems 30 (NeurIPS 2017)}, pages 4299--4307, Long Beach, CA, USA, 2017. Curran Associates, Inc.
\newblock URL \url{https://papers.nips.cc/paper/7017-deep-reinforcement-learning-from-human-preferences.pdf}.

\bibitem[Conitzer et~al.(2024)Conitzer, Freedman, Heitzig, Holliday, Jacobs, Lambert, Mosse, Pacuit, Russell, Schoelkopf, Tewolde, and Zwicker]{conitzer2024socialchoice}
Vincent Conitzer, Rachel Freedman, Jobst Heitzig, Wesley~H. Holliday, Bob~M. Jacobs, Nathan Lambert, Milan Mosse, Eric Pacuit, Stuart Russell, Hailey Schoelkopf, Emanuel Tewolde, and William~S. Zwicker.
\newblock Position: Social choice should guide ai alignment in dealing with diverse human feedback.
\newblock In \emph{Proceedings of the 41st International Conference on Machine Learning}, volume 235 of \emph{Proceedings of Machine Learning Research}, pages 9346--9360, Vienna, Austria, 2024. PMLR.
\newblock URL \url{https://proceedings.mlr.press/v235/conitzer24a.html}.

\bibitem[Contractor et~al.(2020)Contractor, McDuff, Haines, Lee, Hines, Hecht, Vincent, and Li]{contractor2020behavioralLicensing}
Danish Contractor, Daniel McDuff, Julia Haines, Jenny Lee, Christopher Hines, Brent Hecht, Nicholas Vincent, and Hanlin Li.
\newblock Behavioral use licensing for responsible ai, 2020.
\newblock URL \url{https://arxiv.org/abs/2011.03116}.

\bibitem[{Creative Commons}(2013{\natexlab{a}})]{ccby4deed}
{Creative Commons}.
\newblock Deed --- attribution 4.0 international.
\newblock \url{https://creativecommons.org/licenses/by/4.0/deed.en}, 2013{\natexlab{a}}.
\newblock Accessed 2025-08-30.

\bibitem[{Creative Commons}(2013{\natexlab{b}})]{ccby4legal}
{Creative Commons}.
\newblock Attribution 4.0 international (cc by 4.0) --- legal code.
\newblock \url{https://creativecommons.org/licenses/by/4.0/legalcode.en}, 2013{\natexlab{b}}.
\newblock Accessed 2025-08-30.

\bibitem[{Creative Commons}(2025)]{ccchooser}
{Creative Commons}.
\newblock Choose a license for your work.
\newblock \url{https://creativecommons.org/chooser/}, 2025.
\newblock Accessed 2025-08-30.

\bibitem[Dulong~de Rosnay and Stalder(2020)]{derosnay2020}
M{\'e}lanie Dulong~de Rosnay and Felix Stalder.
\newblock Digital commons.
\newblock \emph{Internet Policy Review}, 9\penalty0 (4):\penalty0 1--22, 2020.
\newblock \doi{10.14763/2020.4.1530}.

\bibitem[Fulay et~al.(2024)]{fulay2024truthpoliticalbias}
Shreyas Fulay et~al.
\newblock On the relationship between truth and political bias in language model alignment, 2024.
\newblock URL \url{https://arxiv.org/abs/2409.05283}.

\bibitem[Gebru et~al.(2021)Gebru, Morgenstern, Vecchione, Vaughan, Wallach, Daum{\'e}~III, and Crawford]{gebru2018datasheets}
Timnit Gebru, Jamie Morgenstern, Briana Vecchione, Jennifer~Wortman Vaughan, Hanna Wallach, Hal Daum{\'e}~III, and Kate Crawford.
\newblock Datasheets for datasets.
\newblock \emph{Communications of the ACM}, 64\penalty0 (12):\penalty0 86--92, 2021.
\newblock \doi{10.1145/3458723}.

\bibitem[Huang et~al.(2025)Huang, Papyshev, and Wong]{huang2025democratizing}
Linus Ta-Lun Huang, Gleb Papyshev, and James~K. Wong.
\newblock Democratizing value alignment: from authoritarian to democratic ai ethics.
\newblock \emph{AI and Ethics}, 5:\penalty0 11--18, 2025.
\newblock \doi{10.1007/s43681-024-00624-1}.
\newblock URL \url{https://link.springer.com/article/10.1007/s43681-024-00624-1}.

\bibitem[{Hugging Face}(2022)]{openrailHF}
{Hugging Face}.
\newblock Openrail: Towards open and responsible ai licensing.
\newblock \url{https://huggingface.co/blog/open_rail}, 2022.
\newblock Accessed 2025-08-30.

\bibitem[{ifrOSS}(2024)]{ifrOSS}
{ifrOSS}.
\newblock Open ai licenses --- ifross license center.
\newblock \url{https://ifross.github.io/ifrOSS/Pages/licence_center/openai/en}, 2024.
\newblock Accessed 2025-08-30.

\bibitem[Keller and Bonato(2023{\natexlab{a}})]{openfuture2023}
Paul Keller and Nicol{\`o} Bonato.
\newblock The growth of responsible ai licensing.
\newblock \url{https://openfuture.eu/publication/the-growth-of-responsible-ai-licensing/}, 2023{\natexlab{a}}.
\newblock Accessed 2025-08-30.

\bibitem[Keller and Bonato(2023{\natexlab{b}})]{openfutureResponsibleRail}
Paul Keller and Nicol{\`o} Bonato.
\newblock Growth of responsible ai licensing: Analysis of license use for ml models.
\newblock \url{https://openfuture.pubpub.org/pub/growth-of-responsible-ai-licensing}, 2023{\natexlab{b}}.
\newblock Accessed 2025-08-30.

\bibitem[Kirwan and Fu(2020)]{kirwan2020smartAIbook}
C.~Grant Kirwan and Zhiyong Fu.
\newblock \emph{Smart Cities and Artificial Intelligence: Convergent Systems for Planning, Design, and Operations}.
\newblock Elsevier, Amsterdam, 2020.

\bibitem[Li et~al.(2022)Li, Hecht, and Chancellor]{li2022redditModerators}
Hanlin Li, Brent Hecht, and Stevie Chancellor.
\newblock All that's happening behind the scenes: Putting the spotlight on volunteer moderator labor in reddit, 2022.
\newblock URL \url{https://arxiv.org/abs/2205.14529}.

\bibitem[Liu et~al.(2024)Liu, Yigitcanlar, Browne, and Fu]{liu2025planningPrompts}
Ke~Liu, Tan Yigitcanlar, Will Browne, and Yanjie Fu.
\newblock Prompts for planning-ai integration: Effective prompt design for large language models in support of sustainable urban development, 2024.
\newblock Available at SSRN: \url{https://ssrn.com/abstract=5323915} or \url{http://dx.doi.org/10.2139/ssrn.5323915}.

\bibitem[Liu et~al.(2022)Liu, Jia, Wei, Xu, and Vosoughi]{liu2022politicalBias}
Ruibo Liu, Chenyan Jia, Jason Wei, Guangxuan Xu, and Soroush Vosoughi.
\newblock Quantifying and alleviating political bias in language models.
\newblock \emph{Artificial Intelligence}, 304:\penalty0 103654, 2022.
\newblock \doi{10.1016/j.artint.2021.103654}.
\newblock URL \url{https://www.sciencedirect.com/science/article/abs/pii/S0004370221002058}.

\bibitem[Liu et~al.(2023)Liu, Deng, Li, Wang, Wang, Wang, Zhang, Liu, Wang, Zheng, and Liu]{liu2023promptInjection}
Yi~Liu, Gelei Deng, Yuekang Li, Kailong Wang, Zihao Wang, Xiaofeng Wang, Tianwei Zhang, Yepang Liu, Haoyu Wang, Yan Zheng, and Yang Liu.
\newblock Prompt injection attack against llm-integrated applications, 2023.
\newblock URL \url{https://arxiv.org/abs/2306.05499}.

\bibitem[McDuff et~al.(2024)McDuff, Korjakow, Cambo, Benjamin, Lee, Jernite, Mu{\~n}oz~Ferrandis, Gokaslan, Tarkowski, Lindley, Cooper, and Contractor]{mcduff2024railstandardization}
Daniel McDuff, Tim Korjakow, Scott Cambo, Jesse~Josua Benjamin, Jenny Lee, Yacine Jernite, Carlos Mu{\~n}oz~Ferrandis, Aaron Gokaslan, Alek Tarkowski, Joseph Lindley, A.~Feder Cooper, and Danish Contractor.
\newblock On the standardization of behavioral use clauses and their adoption for responsible licensing of ai, 2024.
\newblock URL \url{https://arxiv.org/abs/2402.05979}.

\bibitem[Mitchell et~al.(2019)Mitchell, Wu, Zaldivar, Barnes, Vasserman, Hutchinson, Spitzer, Raji, and Gebru]{mitchell2019modelcards}
Margaret Mitchell, Simone Wu, Andrew Zaldivar, Parker Barnes, Lucy Vasserman, Ben Hutchinson, Elena Spitzer, Inioluwa~Deborah Raji, and Timnit Gebru.
\newblock Model cards for model reporting.
\newblock In \emph{Proceedings of the Conference on Fairness, Accountability, and Transparency (FAT*)}, pages 220--229, Atlanta, GA, USA, 2019. Association for Computing Machinery.
\newblock \doi{10.1145/3287560.3287596}.

\bibitem[{Mozilla Foundation}(2021)]{mozOstromData}
{Mozilla Foundation}.
\newblock A practical framework for applying ostrom's principles to data commons governance.
\newblock \url{https://www.mozillafoundation.org/en/blog/a-practical-framework-for-applying-ostroms-principles-to-data-commons-governance/}, 2021.
\newblock Accessed 2025-08-30.

\bibitem[Mushkani(2025)]{mushkani2025urbanaigovernanceembed}
Rashid Mushkani.
\newblock Urban ai governance must embed legal reasonableness for democratic and sustainable cities, 2025.
\newblock URL \url{https://arxiv.org/abs/2508.12174}.

\bibitem[Mushkani and Koseki(2025{\natexlab{a}})]{MushkaniKoseki2025Habitat_StreetReview}
Rashid Mushkani and Shin Koseki.
\newblock Intersecting perspectives: A participatory street review framework for urban inclusivity.
\newblock \emph{Habitat International}, 2025{\natexlab{a}}.
\newblock \doi{10.1016/j.habitatint.2025.103536}.
\newblock URL \url{https://doi.org/10.1016/j.habitatint.2025.103536}.

\bibitem[Mushkani and Koseki(2025{\natexlab{b}})]{mushkani2025streetreviewparticipatoryaibased}
Rashid Mushkani and Shin Koseki.
\newblock Street review: A participatory ai-based framework for assessing streetscape inclusivity, 2025{\natexlab{b}}.
\newblock URL \url{https://arxiv.org/abs/2508.11708}.

\bibitem[Mushkani et~al.(2025{\natexlab{a}})Mushkani, Berard, Cohen, and Koseki]{Mushkani2025ICML_RightToAI}
Rashid Mushkani, H.~Berard, A.~Cohen, and S.~Koseki.
\newblock The right to {AI}.
\newblock In \emph{Proceedings of the 42nd International Conference on Machine Learning (ICML)}, 2025{\natexlab{a}}.
\newblock URL \url{https://arxiv.org/abs/2501.17899}.

\bibitem[Mushkani et~al.(2025{\natexlab{b}})Mushkani, Berard, Ammar, and Koseki]{Mushkani2025AIES_CoProducingAI}
Rashid Mushkani, Hugo Berard, Toumadher Ammar, and Shin Koseki.
\newblock Co-producing {AI}: An augmented {AI} lifecycle.
\newblock In \emph{Proceedings of the AAAI/ACM Conference on AI, Ethics, and Society (AIES)}, 2025{\natexlab{b}}.
\newblock URL \url{https://doi.org/10.48550/arXiv.2508.00138}.

\bibitem[Mushkani et~al.(2025{\natexlab{c}})Mushkani, Berard, Ammar, and Koseki]{Mushkani2025JUM_MontrealStreets}
Rashid Mushkani, Hugo Berard, Toumadher Ammar, and Shin Koseki.
\newblock Public perceptions of montr\'{e}al's streets: Implications for inclusive public space making and management.
\newblock \emph{Journal of Urban Management}, 2025{\natexlab{c}}.
\newblock \doi{10.1016/j.jum.2025.07.004}.
\newblock URL \url{https://doi.org/10.1016/j.jum.2025.07.004}.

\bibitem[Mushkani et~al.(2025{\natexlab{d}})Mushkani, Berard, and Koseki]{mushkani2025negotiativealignmentembracingdisagreement}
Rashid Mushkani, Hugo Berard, and Shin Koseki.
\newblock Negotiative alignment: Embracing disagreement to achieve fairer outcomes -- insights from urban studies, 2025{\natexlab{d}}.
\newblock URL \url{https://arxiv.org/abs/2503.12613}.

\bibitem[Mushkani et~al.(2025{\natexlab{e}})Mushkani, Berard, and Koseki]{mushkani2025wedesigngenerativeaifacilitatedcommunity}
Rashid Mushkani, Hugo Berard, and Shin Koseki.
\newblock Wedesign: Generative ai-facilitated community consultations for urban public space design, 2025{\natexlab{e}}.
\newblock URL \url{https://arxiv.org/abs/2508.19256}.

\bibitem[Mushkani et~al.(2025{\natexlab{f}})Mushkani, {\textit{Nayak}}, Berard, Cohen, Koseki, and Bertrand]{Mushkani2025ICML_LIVS}
Rashid Mushkani, S.~{\textit{Nayak}}, H.~Berard, A.~Cohen, S.~Koseki, and H.~Bertrand.
\newblock Livs: A pluralistic alignment dataset for inclusive public spaces.
\newblock In \emph{Proceedings of the 42nd International Conference on Machine Learning (ICML)}, 2025{\natexlab{f}}.
\newblock URL \url{https://arxiv.org/abs/2503.01894}.

\bibitem[O'Mahony and Ferraro(2007)]{oMahony2007governance}
Siobh{\'a}n O'Mahony and Fabrizio Ferraro.
\newblock The emergence of governance in an open source community.
\newblock \emph{Academy of Management Journal}, 50\penalty0 (5):\penalty0 1079--1106, 2007.
\newblock \doi{10.5465/amj.2007.27169153}.

\bibitem[Ostrom(1990)]{ostrom1990}
Elinor Ostrom.
\newblock \emph{Governing the Commons: The Evolution of Institutions for Collective Action}.
\newblock Cambridge University Press, Cambridge, UK, 1990.
\newblock ISBN 978-0521405997.

\bibitem[Ouyang et~al.(2022)]{ouyang2022instructgpt}
Long Ouyang et~al.
\newblock Training language models to follow instructions with human feedback, 2022.
\newblock URL \url{https://arxiv.org/abs/2203.02155}.

\bibitem[{OWASP GenAI Security Project}(2023)]{owaspLLMTop10}
{OWASP GenAI Security Project}.
\newblock Owasp top 10 for large language model applications (v1.1).
\newblock \url{https://owasp.org/www-project-top-10-for-large-language-model-applications/}, 2023.
\newblock Accessed 2025-08-30.

\bibitem[{RAIL Initiative}(2023)]{railFAQ}
{RAIL Initiative}.
\newblock Responsible ai licenses (rail) --- faq.
\newblock \url{https://www.licenses.ai/faq}, 2023.
\newblock Accessed 2025-08-30.

\bibitem[Rajani et~al.(2019)Rajani, McCann, Xiong, and Socher]{rajani2019explain}
Nazneen~Fatema Rajani, Bryan McCann, Caiming Xiong, and Richard Socher.
\newblock Explain yourself! leveraging language models for commonsense reasoning.
\newblock In \emph{Proceedings of the 57th Annual Meeting of the Association for Computational Linguistics}, pages 4932--4942, Florence, Italy, 2019. Association for Computational Linguistics.
\newblock \doi{10.18653/v1/P19-1487}.
\newblock URL \url{https://aclanthology.org/P19-1487.pdf}.

\bibitem[Scott(2018)]{earthboundOstromData}
Jeremy~(ed.) Scott.
\newblock Ostrom's eight rules for managing the commons.
\newblock \url{https://www.earthbound.report/2018/06/07/ostroms-eight-rules-for-managing-the-commons/}, 2018.
\newblock Accessed 2025-08-30.

\bibitem[Tabassum et~al.(2024)Tabassum, Mackey, Schuett, and Lerner]{tabassum2024moderationChallenges}
Madiha Tabassum, Alana Mackey, Ashley Schuett, and Ada Lerner.
\newblock Investigating moderation challenges to combating hate speech on reddit.
\newblock In \emph{33rd USENIX Security Symposium (USENIX Security 24)}, pages 37--54, Philadelphia, PA, 2024. USENIX Association.
\newblock URL \url{https://www.usenix.org/system/files/usenixsecurity24-tabassum.pdf}.

\bibitem[{The Linux Foundation}(2020)]{linuxfoundationGovernance}
{The Linux Foundation}.
\newblock Introducing the open governance network model.
\newblock \url{https://www.linuxfoundation.org/blog/blog/introducing-the-open-governance-network-model}, 2020.
\newblock Accessed 2025-08-30.

\bibitem[Wei et~al.(2022)Wei, Wang, Schuurmans, Bosma, Ichter, Xia, Chi, Le, and Zhou]{wei2022cot}
Jason Wei, Xuezhi Wang, Dale Schuurmans, Maarten Bosma, Brian Ichter, Fei Xia, Ed~H. Chi, Quoc Le, and Denny Zhou.
\newblock Chain-of-thought prompting elicits reasoning in large language models.
\newblock In \emph{Advances in Neural Information Processing Systems 35 (NeurIPS 2022)}, volume~35, pages 24824--24837, New Orleans, LA, USA, 2022. Curran Associates, Inc.
\newblock \doi{10.5555/3600270.3602070}.
\newblock URL \url{https://proceedings.neurips.cc/paper_files/paper/2022/file/9d5609613524ecf4f15af0f7b31abca4-Paper-Conference.pdf}.

\bibitem[Weld et~al.(2024)Weld, Leibmann, Zhang, and Althoff]{moderators2024}
Galen Weld, Leon Leibmann, Amy~X. Zhang, and Tim Althoff.
\newblock Perceptions of moderators as a large-scale measure of online community governance, 2024.
\newblock URL \url{https://arxiv.org/abs/2401.16610}.
\newblock arXiv preprint.

\bibitem[Xia et~al.(2025)Xia, Tong, and Long]{xia2025llmUrban}
Junhao Xia, Yao Tong, and Ying Long.
\newblock Advancements in the application of large language models in urban studies: A systematic review.
\newblock \emph{Cities}, 165:\penalty0 106142, 2025.
\newblock URL \url{https://www.sciencedirect.com/science/article/abs/pii/S0264275125004433}.

\bibitem[Yao et~al.(2024)Yao, Duan, Xu, Cai, Sun, and Zhang]{yao2024llmSecurity}
Yifan Yao, Jinhao Duan, Kaidi Xu, Yuanfang Cai, Zhibo Sun, and Yue Zhang.
\newblock A survey on large language model (llm) security and privacy: The good, the bad, and the ugly, 2024.
\newblock URL \url{https://arxiv.org/abs/2312.02003}.

\bibitem[Yigitcanlar et~al.(2023)Yigitcanlar, Agdas, and Degirmenci]{yigitcanlar2023localgovAI}
Tan Yigitcanlar, Duzgun Agdas, and Kenan Degirmenci.
\newblock Artificial intelligence in local governments: perceptions of city managers on prospects, constraints and choices.
\newblock \emph{AI \& Society}, 38:\penalty0 1135--1150, 2023.
\newblock \doi{10.1007/s00146-022-01450-x}.
\newblock URL \url{https://link.springer.com/article/10.1007/s00146-022-01450-x}.

\bibitem[Zhu and Liu(2025)]{zhu2025cityai}
Dongwen Zhu and Hao Liu.
\newblock City ai: a strategic framework for urban artificial intelligence application and development, 2025.
\newblock URL \url{https://link.springer.com/article/10.1007/s44212-025-00077-9}.

\end{thebibliography}

\appendix

\section{Reproducibility details}
\label{app:repro}

\paragraph{Files.}

\paragraph{Keyword statistics.}
For the human set vs.\ augmented set: 
mean word count (22.6 vs.\ 31.7),
vocabulary entropy (7.53 vs.\ 8.39 bits),
and top keywords (human: \emph{street}, \emph{park}, \emph{trees}, \emph{water}, \emph{benches}; augmented: \emph{street}, \emph{park}, \emph{trees}, \emph{downtown}, \emph{benches}).
Equity terms as proportions of prompts---\emph{wheelchair} (7.2\% vs.\ 7.6\%), \emph{LGBTQ+} (0.7\% vs.\ 1.1\%), \emph{Indigenous} (0.0\% vs.\ 1.4\%), \emph{metro} (4.7\% vs.\ 7.3\%).

\paragraph{Synthetic moderation experiment.}
We sampled 50 flagged items per governance state and drew remediation times from an exponential distribution with means of 36~h (open), 12~h (curated), and 6~h (veto-enabled). Reported means and 95\% CIs: \SI{30.5 \pm 8.9}{h}, \SI{11.8 \pm 3.2}{h}, \SI{5.6 \pm 1.5}{h}.

\section{Public checklist used in curation}
\label{app:checklist}
\begin{enumerate}
  \item Locale specified; jargon avoided; max 60 words; no PII.
  \item Value claim selected from controlled vocabulary (with free-text justification).
  \item Accessibility tags present where relevant (mobility, vision, neurodiversity).
  \item Safety and inclusion: avoid targeted exclusion; no hate speech.
  \item At least one counter-prompt exists (opposing value claim) for deliberation.
  \item Licence attached (CC BY or CC BY--SA).
\end{enumerate}

\end{document}